\documentclass[prb, twocolumn,amsmath,amssymb, superscriptaddress,longbibliography, nofootinbib]{revtex4-1}
\usepackage[pdftex,colorlinks=true]{hyperref}
\hypersetup{
	citecolor = black
}
\usepackage{graphicx}
\usepackage{natbib,hyperref}
\usepackage{amsmath,epsfig,color,amssymb}
\usepackage{bbold}
\usepackage[dvipsnames]{xcolor}
\usepackage[export]{adjustbox}
\usepackage{comment}
\usepackage[titletoc,title]{appendix}
\newcommand{\beg}{\begin{equation}}
\newcommand{\en}{\end{equation}}

\newcommand \bel  {\begin{align}}
\newcommand \enl  {\end{align}}

\newcommand{\ket}[1]{|#1\rangle}

\definecolor{new}{rgb}{.08,.05,.8}

\begin{document}

\author{Didier Ndengeyintwali}
\affiliation{Physics Department, City College of the City University of New York, NY 10031, U.S.A.}
\affiliation{Physics Program, Graduate Center of City University of New York, NY 10031, U.S.A.}

\author{Shiva Heidari}
\affiliation{Physics Department, City College of the City University of New York, NY 10031, U.S.A.}

\author{Cody Youmans}
\affiliation{Physics Department, City College of the City University of New York, NY 10031, U.S.A.}

\author{Pavan Hosur}
\affiliation{Department of Physics, University of Houston, Houston, TX 77204, U.S.A.}
\affiliation{Texas Center for Superconductivity, University of Houston, Houston, TX 77204}

\author{Pouyan Ghaemi}
\affiliation{Physics Department, City College of the City University of New York, NY 10031, U.S.A.}
\affiliation{Physics Program, Graduate Center of City University of New York, NY 10031, U.S.A.}

\title{Anomalous Yu-Shiba-Rusinov spectrum and superconductivity induced magnetic interactions in materials with topological band inversion}

\begin{abstract}
	We study the Yu-Shiba-Rusinov states in materials with bulk band inversion such as iron-based topological superconductors or doped topological insulators. We show that the structure of the YSR state spectrum depends on the doping level relative to the chemical potential at which the band-inversion occurs. Moreover, we demonstrate that the transition from ferromagnetic to antiferromagnetic coupling and vice versa, which is caused by the coupling of magnetic impurities through the overlap of YSR states, is highly dependent on the doping level. Additionally, topological edge states may have a substantial impact on the YSR states, leading to a decrease in YSR state energies and the creation of new states when the magnetic impurity approaches the boundary. 
\end{abstract}

\maketitle

\section{introduction} 
The observation of superconductivity in doped topological insulators \cite{PhysRevLett.104.057001}, as well as the  prediction of possible topological superconductivity\cite{fu2010odd}, have triggered extensive investigations on superconducting doped topological insulators. In recent years, the iron-based superconductor FeSe$_{1-x}$Te$_x$ (FST), which shares similarities in its band structure with doped topological insulator, emerged as a promising candidate to realize topological superconductivity ~\cite{wang2015topological}. It was theoretically predicted that similar to superconducting doped topological insulators \cite{PhysRevLett.107.097001}, the band inversion in the electronic structure of FST leads to the localized Majorana zero-modes at the end of the superconducting vortex~\cite{xu2016topological}. The experimental realisation of vortex zero modes in FST\cite{zhang2018observation,wang2018evidence,zhu2020nearly,machida2019zero,miao2018universal} led to extensive studies. On another front, it was shown that the interplay between the bulk and surface superconducting phases in doped topological insulators leads to novel phenomena such as vortex phase transitions \cite{PhysRevLett.107.097001,ghazaryan2020effect,rui} which emerges due to the electronic band inversion\cite{PhysRevB.76.045302} in the bulk bands. The latter results indicated that the underlying connection of edge states and the bulk band structure would affect the properties of defect modes in the superconducting phase. 

\begin{figure*}[t]
	\centering
	\includegraphics[scale=0.5]{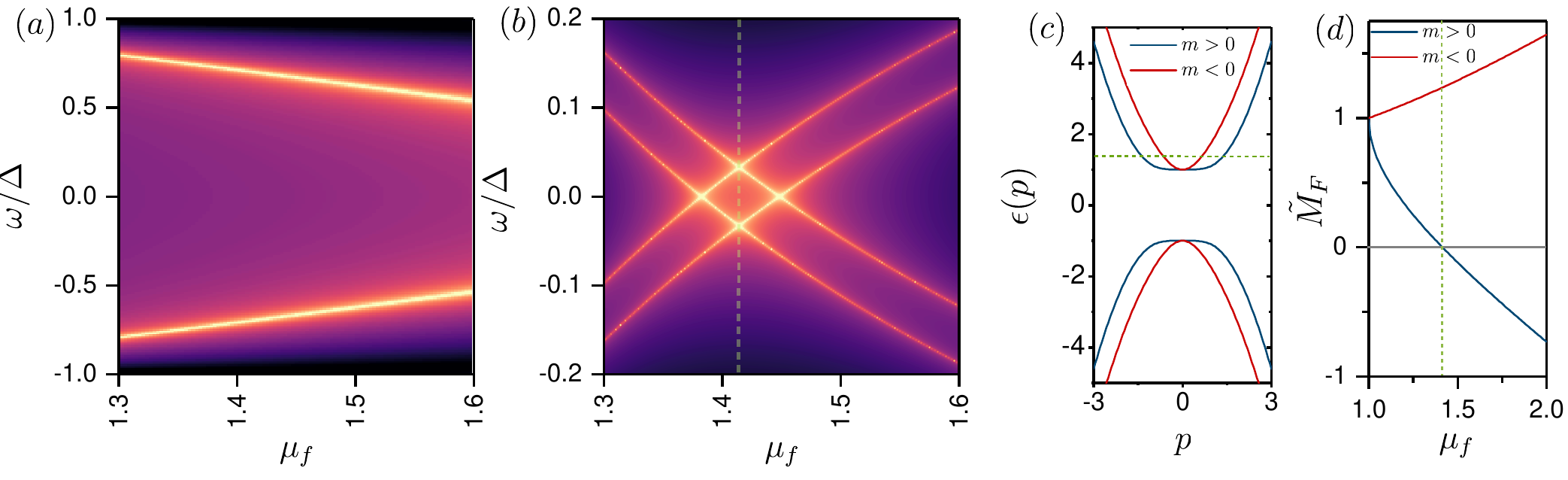} \caption{Single-impurity YSR bound states as a function of chemical potential, for (a): non-topological case ($m<0$) and (b): topological case ($m>0$). The critical chemical potential ($\mu_f^c=\sqrt{M/m}$) is indicated by the dashed line in the topological cases. We set $M=1$, $|m|=0.5$, and ${\cal J}S=20/\pi$. (c): The low-energy bulk bands for the topological (blue), and non-topological (red) cases, respectively. The green dashed line in each figure represents the critical chemical potential where (d): the topological mass $\tilde{M}_F$ changes sign.}
	\label{fig1}
\end{figure*}

Another type of in-gap states in superconductors is Yu-Shiba-Rusinov (YSR) states which can form on the magnetic impurities \cite{shiba1,shiba2,shiba3}.  Iron atoms are common magnetic impurities in FST \cite{PhysRevLett.108.107002,Yin2015} and their corresponding YSR states are observed \cite{yin2015observation,fan2021observation,PhysRevLett.120.156803}. The YSR states in FST have particular features, such as their robust appearance at zero energy, which are not common in other superconducting materials. Different theoretical models have attributed the anomalous features of YSR states to phenomena such as the development of anomalous vortices at magnetic impurity sites \cite{PhysRevX.9.011033}, interaction between modes in different vortices \cite{chiu2020scalable}, the effect of topological band structure in the parent insulating phases \cite{PhysRevLett.128.237001} and modification of the form of superconducting pairing as a result of the Zeeman field of magnetic impurities\cite{PhysRevB.106.L201107}. On another front, interaction between magnetic impurities arranged on a line in a superconductor, can lead to formation of one dimensional topological superconductors with the potential for generating Majorna states \cite{nadj2014observation,PhysRevB.97.125119}.Interaction between magnetic impurities in superconductors results from overlap between YSR states localized on different impurities. It has been shown that the nature of the superconducting state, as well as the presence of spin-orbit coupling \cite{PhysRevB.91.064505,PhysRevLett.114.236804,Park_2023}, could affect the interaction of magnetic impurities induced by the YSR states. The possibility of YSR states going through quantum phase transition has been also of particular current interest\cite{PhysRevB.89.180505,PhysRevB.106.115409}.

In this paper we study the YSR states in superconducting doped topological insulators. Our efforts are motivated by previous results on the effects of doping level on vortex modes in doped topological insulators and resulting vortex phase transition. We show that the presence of band inversion in the bulk bands has a crucial effect on the properties of YSR states and coupling between magnetic impurities which is induced through overlap of YSR states in superconducting phase. In particular, anomalously strong dependence of magnetic coupling on the doping level and local charge fluctuations arises due to the band inversion in the bulk bands. The topological edge states near the boundaries, even though they are gapped by the superconductivity, push the YSR states closer toward the middle of superconducting gap. The emergence of zero energy YSR states has drawn a lot of attention since they may shed light on the emergence of Majorana zero mode, which is promising for realizing fault-tolerant quantum computing\cite{Sarma2015}.

The rest of the paper is organized as follows: In section \ref{mh} we present the bulk model Hamiltonian and the resulting YSR spectrum of magnetic impurities. In section \ref{cop} we examine the coupling of magnetic impurities induced by overlap of YSR states and its dependence on the chemical potential. In section \ref{cd} we examine the enhanced effect of charge disorder on the magnetic coupling. In section \ref{FAF} we show that the presence of bulk band inversion leads to strong dependence of the type of magnetic coupling on the chemical potential and leads to transition between ferro and antiferromagnetic coupling by tuning the chemical potential.

\section{Model Hamiltonian}\label{mh}
In this paper, we use the following four band tight-binding Hamiltonian in a cubic lattice corresponding to a $Z_2$ topological insulator \cite{PhysRevB.81.045120},
\begin{equation}\label{modelH}
	H
	_L(\textbf{K})=v_p (sin(K_x) \gamma^1+sin (K_y) \gamma^2+sin(K_z) \gamma^3)+M(\textbf{K}) \gamma^5
\end{equation}
where $M(\textbf{K})=-M_0+M_1 (\cos(K_x)+\cos(K_y))+M_2 \cos(K_z)$. We have defined $\gamma^1=\tau^1 \otimes \sigma^1$, $\gamma^2=\tau^1 \otimes \sigma^2$, $\gamma^3=\tau^1 \otimes \sigma^3$, and $\gamma^5=\tau^3 \otimes \sigma^0$, where the Pauli matrices $\sigma^i$ and $\tau^i$ act on the spin and orbital degrees of freedom, respectively. The model Hamiltonian in Eq.~\ref{modelH} is directly relevant for two orbital topological insulators, e.g., Bi$_2$Se$_3$, whereas for the iron-based superconductors with band inversion, such as FeSe$_{x}$Te$_{1-x}$, the effective Hamiltonian includes larger number of orbitals \cite{wang2015topological,cvetkovic2013space,PhysRevB.106.L201107}. With the parameters $v_p=1$, $M_0=2$, $M_1=1$ and $M_2=-1$, $H_L(p)$ has a single band inversion at $Z=(0,0,\pi)$ between energy bands with opposite parities which describes a $Z_2$ topological band \cite{PhysRevB.76.045302}. Therefore, with this choice of parameters the surface Dirac cone emerges on the edge. The band inversion in the bulk electronic bands, which is captured by the Hamiltonian in Eq.~\ref{modelH}, is the main origin of the phenomena discussed in this paper. Given the presence of band inversion close to the Fermi energy in FeSe$_{x}$Te$_{1-x}$, we expect similar phenomena to emerge in the latter material as well. The effective bulk Hamiltonian for the 3D topological insulator near the band inversion point, i.e., $Z=(0,0,\pi)$, in the continuum limit is given by the following continuum Hamiltonian 
\begin{equation}
	H(\textbf{p})= v_f\  \textbf{p}\cdot \mathbf{\gamma}+ \tilde{M}(p) \gamma^5,
\end{equation} \label{Hp}
where $\mathbf{\gamma}=(\gamma^1,\gamma^2,\gamma^3)$, $\textbf{p}$ is the momentum relative to the $Z=(0,0,\pi)$ point and $p=|\textbf{p}|$. $\tilde{M}(p)=(M-m p^2)$ with $M=3M_1-M_0$, and $m=M_1/2$ is the momentum dependent mass gap. In the rest of the paper we set $v_f=1$, without loss of generality. We note that $2 M$ is the energy gap between valence and conduction bands and is considered positive in our model. The topological characteristic can be determined by the sign of the parameter $m$ as sign$(m)>0$ leads to band inversion; otherwise, the band gap is topologically trivial [Fig.~\ref{fig1}(c)].

Similar to the previous studies on the vortex modes spectrum in doped topological insulators \cite{PhysRevLett.107.097001,ghazaryan2020effect}, we consider intra-orbital singlet superconducting pairing in the bulk, corresponding to the following Bogoliubov-De Gennes (BdG) Hamiltonian:

\begin{equation}
	H_{\text{BdG}}=\mu^3 (H(\textbf{p})-\mu_f \gamma^0)+\mu^2 \gamma^4\Delta.
\end{equation}

Here, $\mu^i$ acts in Nambu particle-hole space,$\gamma^0=\tau^0 \otimes \sigma^0$, $\gamma^4=\tau^0 \otimes \sigma^2$, and the s-wave pairing potential is $\Delta$. The chemical potential $\mu_f$ can be tuned into the valence or conduction bands with significant impacts on the YSR bound-state energies near the impurity. The YSR states result from adding magnetic interaction with the impurity spin of the form $V^{\text{imp}}_{i}(\textbf{r})=-{\cal J}  \sum_i \textbf{S}_{i} \cdot \sigma \ \delta(\textbf{r}-\textbf{r}_{i})$ to the Hamiltonian, where the spin of the impurity is defined as $\boldsymbol{S}=S \hat{z}$, ${\cal J}$ determines the  coupling strength between impurities and the electrons and the index $i$ runs over all the impurities at the position $r_i$. This impurity potential corresponds to impurities with large spins which can be treated classically in the limit $S\rightarrow \infty$, while simultaneously ${\cal J} \rightarrow 0$, so that ${\cal J}S=$const \cite{balatsky2006impurity}. In this regime, we ignore the Kondo scattering effect and the localized spin acts as a local magnetic field. We only consider the purely magnetic scattering term.
\\\\
\section{The YSR spectrum}
To determine the YSR energy spectrum and the wave functions, we start with the following Schr\"{o}dinger equation:
\begin{equation} \label{Sch}
	[\omega-H_{\text{BDG}}] \psi_{\textbf{p}}=\sum_j V_j^{\text{imp}} e^{-i \bf{p}\cdot \bf{r}_j} \psi(\bf{r}_j),
\end{equation}
where $\omega$ is the energy of the YSR states and the sum is over  impurity potentials. Eq.~\ref{Sch} can be written in terms of Green's function (GF) in the following form\cite{GF1,GF2,GF3,GF4,GF5,GF6,GF7} 
\begin{equation}\label{gf}
	\psi(\textbf{r}_i)=\sum_j V_j^{\text{imp}} G(\omega,\textbf{r}_{ij}) \psi(\textbf{r}_j),
\end{equation}
where $\textbf{r}_{ij}=\textbf{r}_i-\textbf{r}_j$ and $G(\omega,\textbf{r}_{ij})=\int \frac{d\textbf{p}}{(2\pi)^3} e^{i \textbf{p}\cdot \textbf{r}_{ij}} G(\omega,\textbf{p})$, is the Fourier transform of the momentum-space GF $G(\omega,\textbf{p})=\left(\omega-H_{\text{BDG}}\right)^{-1}$. The spectrum of the YSR states is then determined by requiring the Eq.~\ref{gf} to have a non-trivial solution. The procedure above can be applied to a single or multiple impurities which incorporates the interaction between the impurities. The GF for the BdG Hamiltonian in momentum space reads as
\begin{equation} \label{GF}
	\begin{split}
		& G(\omega,\textbf{p})=\left(\omega-H_{\text{BDG}}\right)^{-1}\\ & =\dfrac{\mu^3 (\eta^2 H_--(\mu_f^2-\lambda^2)H_+)+\hat{\Delta} (\eta^2+H_
			+^{2})}{(\eta^2+\epsilon_+^2)(\eta^2+\epsilon_-^2)},
	\end{split}
\end{equation}
where $\hat{\Delta}=\omega\mu^0+\Delta \mu^1$, $H_{\pm}=H(\textbf{p})\pm \mu_f$, $\lambda=\sqrt{ p^2+\tilde{M}(p)^2}$, $\epsilon_\pm=-\mu_f \pm \lambda$ and $\eta=
\sqrt{\Delta^2-\omega^2}$.
The full GF in Eq. \ref{GF} can be written as the sum of ${\cal G}_+(\omega,\textbf{p})$ and ${\cal G}_-(\omega,\textbf{p})$, where
\begin{equation}
	{\cal G}_{\pm}(\omega,\textbf{p})=\dfrac{1}{2 \lambda} \dfrac{\mu^3 \epsilon_{\pm}+\Delta \mu^1+\omega}{\epsilon_{\pm}^2+\Delta^2-\omega^2} (\lambda \gamma^0\pm \tilde{M}(p) \gamma^5 \pm  \textbf{p}\cdot \gamma),
\end{equation}
correspond to the GFs where the contribution of the conduction or valence bands are dominant, i.e., $\mu_F> M$ or $\mu_f<M$, respectively.

\begin{figure}[htp]\label{impj}
	\centering
	\includegraphics[width=\columnwidth]{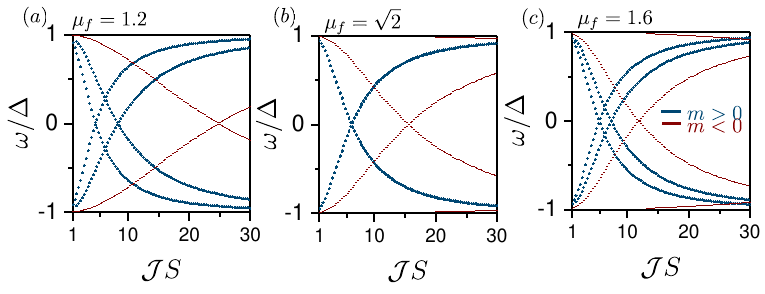} \caption{The single impurity-induced YSR bound state energies as a function of impurity potential strength ${\cal J}S$, at the critical chemical potential $\mu_f^c=\sqrt{2}$, upper the $\mu_f^c$, i.e, $\mu_f=1.6$, and lower than $\mu_f^c$, i.e, $\mu_f=1.2$. The blue and red lines represent the topological and trivial phases. We set the parameters $M=1$, and $m=0.5$ for the topological case and $m=-0.5$ for the non-topological case. }
	\label{fig2}
\end{figure}

Using the real space GF at the origin ($\bf{r}=0$) [Appendix. \ref{appa}],
and assuming that $\mu_f>M$, the spectrum of YSR states associated with a single impurity, is determined by the roots of the equation
$\text{det}(\textbf{1}\pm {\cal J}S G_0(\omega))=0$, in which $G_0(\omega)$ is the local single impurity GF describe as
\begin{equation}
	G_0(\omega)=\int \frac{d \textbf{p}}{(2\pi)^3} {\cal G}_+(\omega,\textbf{p})=\dfrac{\pi \nu_0}{2 \eta}  (\omega+\Delta \mu^1)(\gamma^0+ \dfrac{\tilde{M}_F }{\mu_f}\gamma^5),
\end{equation}
where $\tilde{M}_F$ is the topological mass term at the Fermi surface, i.e.,  $\tilde{M}_F=\tilde{M}(\textbf{p}=\textbf{p}_F)$, and $\nu_0=\frac{p_f \mu_f}{2 \pi^2  (1-2m \tilde{M}_F)}$ is the density of states. The resulting spectrum of YSR states is given by
\begin{equation} \label{singl_bound}
	\omega_{\pm \pm}=\pm \Delta \dfrac{1-\beta_\pm^2}{1+\beta_\pm^2},
\end{equation}
where $\beta_{\pm}=\tilde{\nu}(1\pm \dfrac{\tilde{M}_F}{\mu_f})$, and $\tilde{\nu}=\dfrac{\pi}{4} {\cal J}S \nu_0$. 
Note that at the critical chemical potential $\mu_f^c=\sqrt{M/m}$ where the topological mass term changes sign, i.e., $\tilde{M}(\textbf{p}=\textbf{p}_f^c)=0$, the two subgap energies cross and lead to a degeneracy, i.e., $\omega_{+,+}=\omega_{+,-}$ and $\omega_{-,+}=\omega_{-,-}$.

Fig.~\ref{fig2} shows the spectrum of YSR states as a function of coupling strength with impurity spin and Fig.~\ref{fig1} shows the spectrum of YSR states as a function of the chemical potential for topological and non-topological doped insulators. The YSR in-gap state energies are expressed in units of the half of the superconducting gap i.e., $\omega/\Delta$ in all figures. We note that the size of the superconducting gap in FTS, which is an example of superconductor with electronic band inversion in normal state, is reported as $2\Delta=2-3$ meV~\cite{zhang2018observation}. It can be seen in the Fig. \ref{fig1}.(b) that YSR state spectrum is strongly correlated with chemical potential. In particular, the gap between the YSR state energies in the particle sector vanishes at the critical chemical potential where $\tilde{M}(\textbf{p})$ vanishes and changes sign in the topological case. The critical chemical potential, which identifies the band inversion point, is represented by the vertical dash line in Fig. \ref{fig1}.(b). The effect of the band inversion on the vortex modes in superconducting doped topological insulators and the associated vortex phase transition was previously studied \cite{PhysRevLett.107.097001}. Experimental observation of  the vortex phase transition is challenging due to the large number of vortex modes and the small energy gap, which is of the order of $\Delta^2/\epsilon_f$, where $\Delta$ is the superconducting gap and $\epsilon_f$ is the Fermi energy. The number of in-gap YSR states is much smaller than the vortex modes. The YSR states splitting gap given in Eq.~\ref{singl_bound}, which is of the order of a superconducting gap, is considerably larger than that of vortex modes \cite{Xia2022,PhysRevLett.117.186801,PhysRevB.103.205424}. Consequently, the evolution of the YSR spectrum with chemical potential and the emergent degeneracy at the 
critical chemical potential are much more promising to be experimentally observed. Experimental determination of the critical chemical potential through the variation of the YSR spectrum by doping level makes it possible to verify the phenomena predicted in this paper. 
\\
\section{Effect of chemical potential on YSR induced coupling of magnetic impurities}\label{cop}
The application of the coupled YSR states for the realization of Majorana modes \cite{nadj2014observation} has motivated extended studies of
the YSR states induced coupling of magnetic moments in a superconductor\cite{PhysRevB.97.125119,PhysRevB.89.180505}. The coupling of magnetic impurities in the superconducting state can be examined through the spectrum of multiple YSR states and its dependence on the separation of the impurities. According to Eq. \ref{gf}, the spectrum of YSR states induced by two magnetic impurities ($i,j=1,2$) with impurity potentials $V_{1(2)}=-{\cal J}S_{1(2)}^z \sigma_z$, corresponds to the non-trivial solutions of the equation below,
\begin{equation} \label{dimer-eq}
	\text{det}\begin{pmatrix}
		\textbf{1}-G_0(\omega) V_1 && -G(\omega,r) V_2 \\
		-G^\dagger (\omega,r) V_1 && \textbf{1}-G_0(\omega) V_2
	\end{pmatrix} =0,
\end{equation}
where $G_0(\omega)$ is the local GF and $G(\omega,r)$ is the non-local two-site propagator GF, in which $\textbf{r}=\textbf{r}_1-\textbf{r}_2$, and $r=|\textbf{r}|$ is the separation of the two impurities. In the absence of $G(\omega,r)$, the diagonal part in Eq. \ref{dimer-eq} results in the YSR states spectrum generated by each individual magnetic impurity. The off-diagonal parts which depend on $G(\omega,r)$ modify the energy of single YSR states by taking into account the effects of the hybridization of YSR states between the impurities at the distance of r. Accordingly, the main effect of interactions is implemented through $G(\omega,r)$:
\begin{equation} \label{GFr}
	\begin{split}
		& G(\omega,r)=\\ &\dfrac{\pi \nu_0}{2} [ ( \dfrac{\hat{\Delta}}{\eta} f_1(q)+\mu^3 f_2(q)) (\gamma_0+\dfrac{\tilde{M}_F}{\mu_f} \gamma^3)\\ & +i \lbrace (\dfrac{\hat{\Delta} \ }{\eta}\dfrac{p_f}{\mu_f}-\dfrac{\eta}{\mu_f} \mu^3) f_3(q)+  (\mu^3  \dfrac{p_f}{\mu_f}+\dfrac{\hat{\Delta}}{\mu_f})f_4 (q)\rbrace \gamma^3 ].
	\end{split}
\end{equation}
Here, $q=r p_F$, and the r-dependent functions are defined as
$f_1(q)=\frac{\sin q}{q}$, $f_2(q)=\frac{\cos q -1}{q}$, $f_3(q)=\frac{\sin q-q \cos q}{q^2}$, and $f_4(q)=\frac{q \sin q+ \cos q-1}{q^2}$. Obviously, by separating two impurities to the long distances $r \rightarrow \infty$, the interacting GF, $G(\omega,r)$ will vanish by $1/r$.  The explicit derivation of equation \ref{GFr} is given in Appendix \ref{appb}.
\begin{figure}[t]
\centering
\includegraphics[width=\columnwidth]{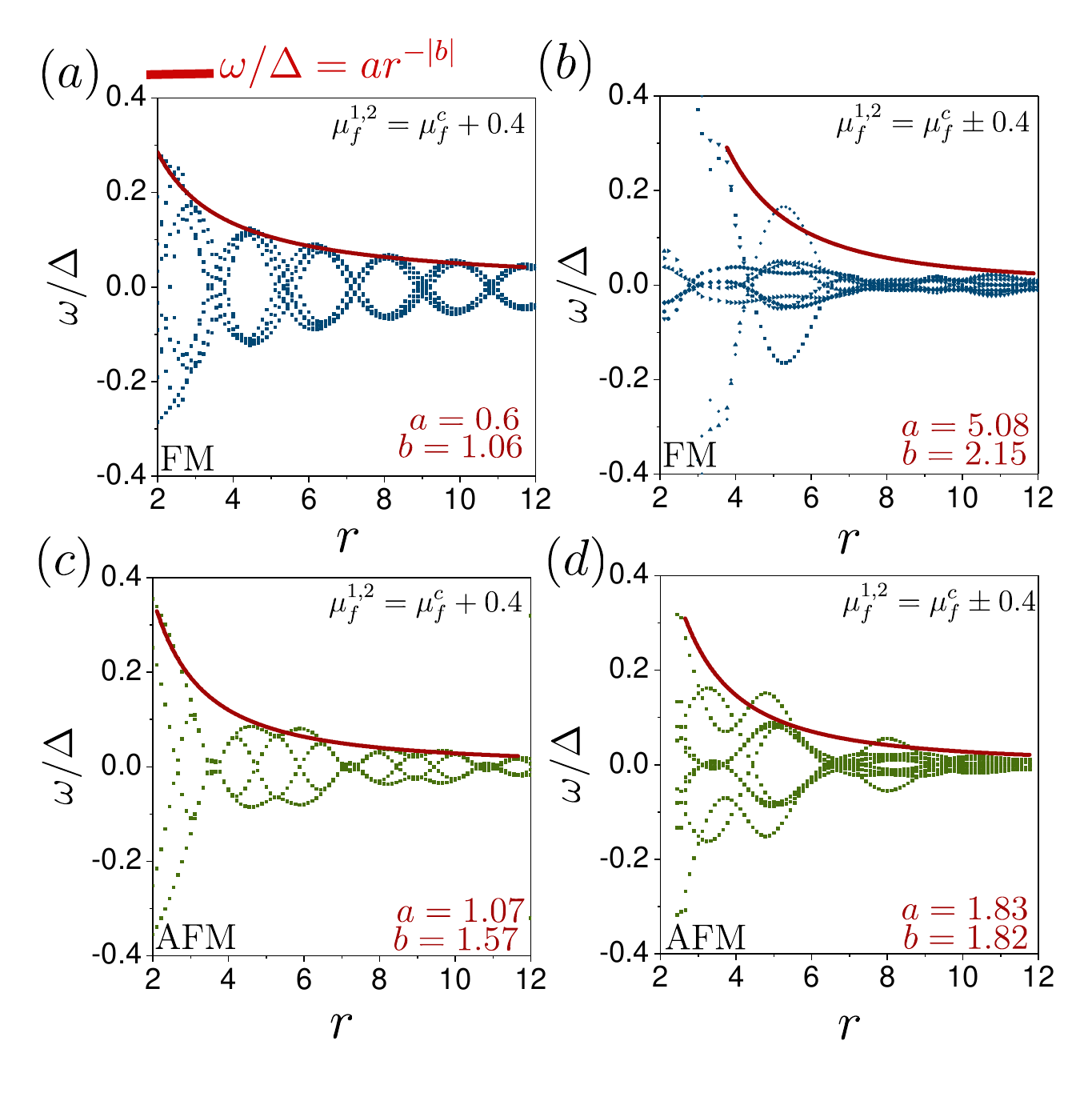} \caption{Change in the  lowest YSR energies of the two impurities with respect to the lowest YSR state of single impurity as a  function of their separation. As the separation increases, the energy of YSR states approaches the energies of YSR states at the single impurity. (a) and (b) correspond to Ferromagnetic and (c) and (d) correspond to antiferromagnetic arrangement In (a) and (c) the chemical potential at the position of the two impurities are the same. In (b) and (d) the chemical potential at the position of the two impurities are above and below critical chemical potential. We set $M=1$, $m=0.5$, $v_f=1$ and ${\cal J}S=20/\pi$. The decay rate is fitted to $\omega/\Delta=a r^{-|b|}$. The dimer exists in a longer range for smaller values of the power $|b|$. In this figure, FM (AFM) stands for (anti-)ferromagnetic.}
\label{fig3}
\end{figure}
Given the dependence of the $G(\omega,r)$ on the mass parameter $\tilde{M}(\textbf{p})$, and the fact that the dominant contribution to the GFs comes from the states close to the Fermi surface, we would expect that the binding energies in the topological case have a dependence on the chemical potential due to the band inversion, i.e., momentum dependence of $\tilde{M}(\textbf{p})$. As we will show below the binding energy of magnetic impurities strongly depends on the chemical potential when the separation of impurities is smaller than the superconducting coherent length.

The binding term of the dimer YSR energies is given by $G(\omega,r) G^\dagger(\omega,r)$ (Eq.~\ref{dimer-eq}), which can be split into two general terms, i.e., $A_1 r^{-a_1}+A_2 r^{-a_2(>-a_1)}$, respectively, where $A_1= \tilde{\nu}^2 (1-(\frac{\tilde{M}_F}{\mu_f})^2)^2= \tilde{\nu}^2 (\frac{ p_f}{\mu_f})^4$, and $A_2 =\tilde{\nu}^2 (\frac{p_f}{\mu_f})^2$. Then, $A_1/A_2= (\frac{p_f}{\mu_f})^2<1$, which can approach to its maximum value at the critical chemical potential. Therefore, the slowly decaying term is suppressed as the chemical potential moves away from the critical chemical potential $\mu_f^c$.

\section{effect of charge disorder on YSR induced magnetic couplings}\label{cd} The presence of charge disorder can naturally lead to the chemical potential fluctuations in the materials. In particular, the chemical potential variation has been reported in FeSe$_x$Te$_{(1-x)}$ \cite{PhysRevResearch.3.L032055,Cho2019}, which has been extensively studied in recent years as an example of a superconductor with band inversion in electronic bands. Interestingly, assuming a short-range form for the impurity potential, the GF directly depends on the chemical potential at the position of the impurity. As a result, the effect of chemical potential variation can be examined using the GF method. The GF form of the Schr\"{o}dinger equation is given by 
\begin{figure}[t]
\centering
\includegraphics[width=\columnwidth]{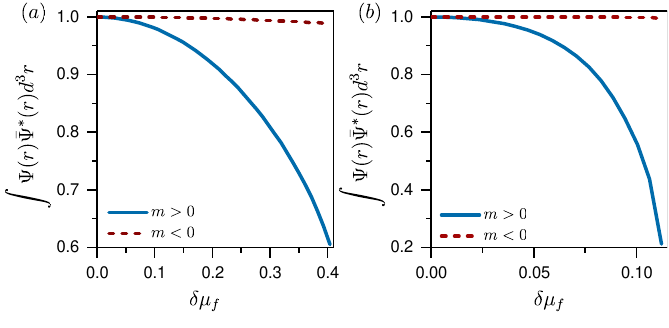} \caption{Overlap integral of two wave functions with  $\mu_f^{1,(2)}=\mu_f^c \pm \delta \mu_f$ where (a): $\mu_f^c=\sqrt{2}$, and $M=1$ (Bottom of the band), $m=0.5$, and (b): $\mu_f^c=0.21$, and $M=0.1$ (Bottom of the band), $m=2.2$.  $\delta \mu_f=0$ indicates that two impurities possess the critical chemical potential. However, when $\delta \mu_f$ approaches to $\delta \mu_f^c=\mu_f^c-M$ corresponding to (a): $0.41$ and (b): $0.11$, means $\mu_f^{(1)}$ comes close to the bottom of the band ($\mu_f^{(1)}=M$) while  $\mu_f^{(2)}$ is in the band ($\mu_f^{(2)}=2\mu_f^c-M$).}
\label{fig4}
\end{figure}
\begin{equation}
\psi(\textbf{r}_i)=\sum_j  V^{\text{imp}}_j G(\omega,\mu_f^j;\textbf{r}_i-\textbf{r}_j) \psi(\textbf{r}_j),
\end{equation}
which leads to
\begin{equation} \label{Eq.dimer}
\begin{pmatrix}
	\textbf{1}-G_0(\omega,\mu_f^{(1)}) V_1 && -G(\omega,\mu_f^{(2)},r) V_2 \\
	-G^\dagger(\omega,\mu_f^{(1)},r) V_1 && \textbf{1}-G_0(\omega,\mu_f^{(2)}) V_2
\end{pmatrix} \begin{pmatrix}
	\psi_1 \\
	\psi_2 
\end{pmatrix}=0.
\end{equation}
The above equation impose the requirement that the determinant of the matrix is zero, which results in a polynomial equation of $\omega$ for each value of chemical potential $\mu_f$. Fig. \ref{fig3} represents the output of the numerical calculations employed to find the roots of the determinant for each $\omega$ and $\mu_f$.
We define the interaction energies of YSR states, as indicated in Fig.~\ref{fig3}, as subtracting the corresponding single impurity-induced YSR energies from the dimer YSR energies. The results demonstrate how coupling of magnetic moments can be significantly impacted by small fluctuations in local chemical potential near critical chemical potential. The line $\omega/\Delta= a r^{-|b|}$ has been fitted to determine the decay rates of the variation in the YSR energies at two impurities due to their hybridizition in the absence (Fig. \ref{fig3}. (a,c)) or presence (Fig. \ref{fig3}. (b,d)) of the chemical potential imbalance.
According to our numerical results shown in Fig .\ref{fig3}, when two impurities have different chemical potentials in a way that one is higher and the other is lower than the critical point $\mu_f^c$, their interaction energy decays much faster compared to when they have the same chemical potential.

The effect of chemical potential fluctuations on YSR state induced coupling of magnetic moments could be understood by examining the overlap of YSR wave functions localized at two impurities. The overlap wave function of two localized YSR states with opposite signs of $\tilde{M}_F$ with $\mu_f=\mu_f^{1}$ at $\textbf{r}_1$ and $\mu_f=\mu_f^{2}$ at $\textbf{r}_2$ is defined as
\begin{equation} \label{overlap}
\int \Psi_{\mu_f^1}^*(\textbf{r}-\textbf{r}_1) \Psi_{\mu_f^2}(\textbf{r}-\textbf{r}_2) d\textbf{r}.
\end{equation}

The detailed derivation of wave-function of YSR states at the position of $\bf{r}$, i.e., $\Psi(\bf{r})$, is given in Appendix.~\ref{appc}.

In the limit of $\textbf{r}_1 \rightarrow \textbf{r}_2$ and using the wave function components in Appendix.\ref{appc}, the overlap integral approaches
\begin{equation}\label{I1}
\begin{split}
	&\int \Psi_{\mu_f^{1}}^*(\textbf{r}-\textbf{r}_1) \Psi_{\mu_f^{2}}(\textbf{r}-\textbf{r}_2) d\textbf{r}=\\ &\dfrac{2 \tilde{\nu}_1 \tilde{\nu}_2}{\sqrt{N_1 N_2}}  \lbrace (1+\dfrac{\tilde{M}_F^{1} \tilde{M}_F^{2}}{\mu_f^{1} \mu_f^{2}}) (\dfrac{1+\omega_1}{\eta_1}\dfrac{1+\omega_2}{\eta_2} {\cal G}_1+{\cal G}_2)+\\ &(\sqrt{1-(\dfrac{\tilde{M}_F^{1}}{\mu_f^{1}})^2}\sqrt{1-(\dfrac{\tilde{M}_F^{2}}{\mu_f^{2}})^2})(\dfrac{1+\omega_1}{\eta_1}\dfrac{1+\omega_2}{\eta_2} {\cal G}_3+{\cal G}_4) \rbrace ,
\end{split}
\end{equation} 
where $\eta_{1(2)}=\sqrt{\Delta^2-\omega_{1(2)}^2}$, ${\cal G}_i=\int f_i^2(q,\alpha) \ r^2 dr d\alpha d \phi$, with $q=r p_f$ and the normalization factor of wave functions with chemical potential $\mu_f^{1(2)}$ is given by
\begin{equation} 
\begin{split}
	&N_{1,2}=2\tilde{\nu}_{1,2} \lbrace (1+(\dfrac{\tilde{M}_F^{1,2}}{\mu_f^{1,2}})^2) ((\dfrac{1+\omega_1}{\eta_1})^2 {\cal G}_1+{\cal G}_2) \\ & +(1-(\dfrac{\tilde{M}_F^{1,2}}{\mu_f^{1,2}})^2) ((\dfrac{1+\omega_{2}}{\eta_{2}})^2 {\cal G}_3+{\cal G}_4)  \rbrace .
\end{split}
\end{equation}
Fig.~\ref{fig4} shows the numerical calculation of the overlap integral (Eq.~\ref{I1}) between two YSR wave functions with $\mu_f^{1,2}=\mu_f^c \pm \delta \mu_f$. 
The real space functions, $f_i(q,\alpha)$, are derived through numerical integration in momentum space (detailed in Appendix \ref{appc}). Subsequently, we perform additional numerical integration over real space to obtain the ${\cal G}_i$ coefficients, which are crucial components in the equation describing the overlap of wave functions [Eq.~\ref{I1}]. Based on the result represented in Fig.~\ref{fig4}, if there are two impurities with the same chemical potential at the critical value, then $\delta \mu_f=0$ or $\mu_f^{1}=\mu_f^{2}=\mu_f^c$. On the other hand, when one of the  chemical potentials approaches the bottom of the band, it leads to $\delta \mu_f=\mu_f^c-M$, i.e., $\mu_f^{1}=M$, and $\mu_f^{2}=2\mu_f^c-M$.
According to the overlap integral in Eq.~\ref{I1}, the overlap of two wave functions can be more suppressed when the critical chemical potential $\mu_f^c$ is close to the bottom of the band and simultaneously $\mu_f^{1}<\mu_f^c$ and $\mu_f^{2}>\mu_f^c$. Fig.~\ref{fig4}.(b) represents the overlap integral for the critical point much closer to the bottom of the band compared to (a). As we can see, there may be more suppression in the overlap wave functions associated with chemical potentials closer to the bottom of the band.
\\
\begin{figure}[t]
\centering
\includegraphics[width=\columnwidth]{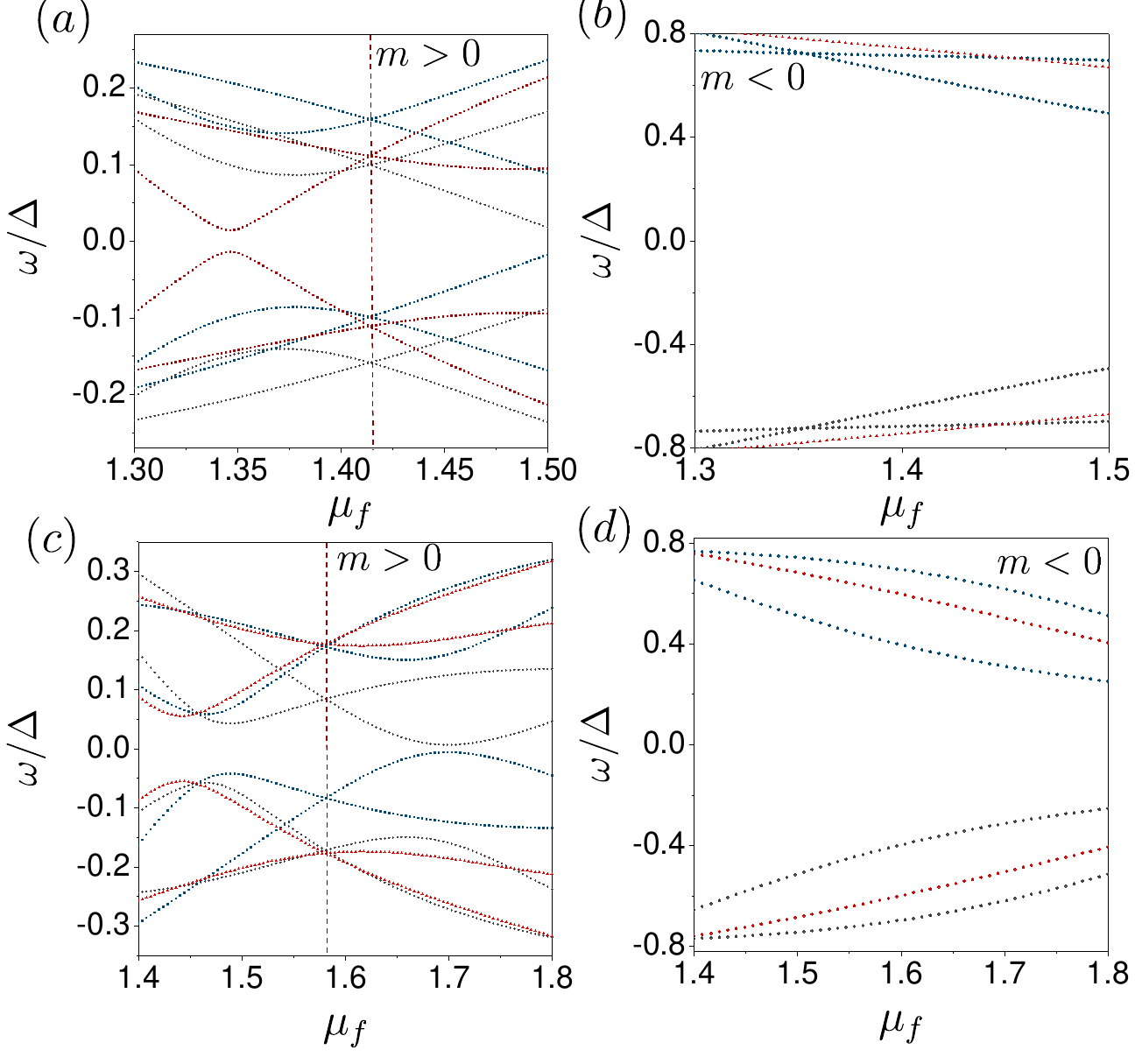} \caption{ YSR bound states near a dimer as a function of chemical potential, for topological case ($m>0$: (a),(c)) and non-topological case ($m<0$: (b),(d)). The critical chemical potential is indicated by the dashed red line. We set $r=5$, ${\cal J}S=20/\pi$, and (a),(b): $M=1$, $m=0.5$ and (c),(d): $M=1$, $m=0.4$. The blue and red lines indicate the ferromagnetic and antiferromagnetic dimer configuration.}
\label{fig5}
\end{figure}
\section{transition between Ferro and anti-ferro coupling of magnetic impurities controlled by chemical potential}\label{FAF}
\begin{figure}[t]\label{impdj}
\centering
\includegraphics[width=\columnwidth]{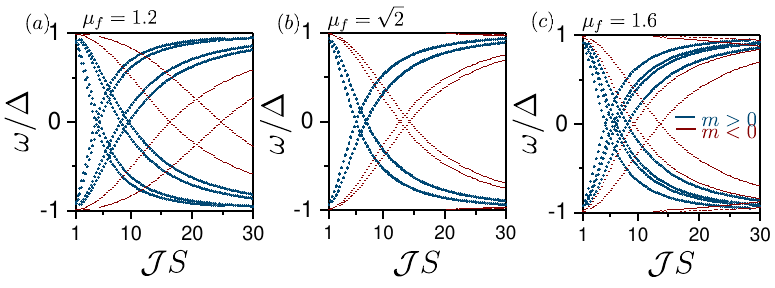} \caption{ The dimer-induced YSR bound state energies as a function of impurity potential strength ${\cal J}S$, at the critical chemical potential $\mu_f^c=\sqrt{2}$, upper the $\mu_f^c$, i.e, $\mu_f=1.6$, and lower than $\mu_f^c$, i.e, $\mu_f=1.2$. The blue and red lines represent the topological and trivial phases. We set the parameters $M=1$, and $m=0.5$ for the topological case and $m=-0.5$ for the non-topological case. }
\label{fig6}
\end{figure}

It has been shown that the magnetic order in a dimer induced by YSR states could be  ferromagnetic or antiferromagnetic \cite{GF2,PhysRevB.90.060401,PhysRevB.91.064505}. In this section, we examine the effect of chemical potential on the type of the magnetic coupling between two impurities. To this end, we derive the binding energy of ferromagnetic and antiferromagnetically oriented impurities. The type of the coupling corresponds to the configuration with lower energy. Fig.~\ref{fig5} shows the YSR state energy induced by two magnetic impurities for both ferromagnetic (blue), i.e., $V_1,V_2>0$, and antiferromagnetic (red), i.e., $V_1>0, V_2<0$ configurations as a function of chemical potential. Fig. \ref{fig6} illustrates the numerical calculations of YSR energy states as a function of impurity potential ${\cal J}S$ for the ferromagnetic case.

In Fig.\ref{fig5}, at the critical chemical potential, the two ferromagnetic and the two anti-ferromagnetic phases are independently degenerate. The degeneracy could be understood from Eq.~\ref{GFr}. At the critical point when $\tilde{M}_F|_{\mu_f=\mu_f^c}=0$, the energy of YSR states corresponding to two eigenvalues of Pauli operator $\tau_3$ (which acts on the orbital space) are degenerate. Therefore, the eigenstates corresponding to the two possible ferromagnetic or antiferromagnetic configurations are independently degenerate. In the GF framework the degeneracy corresponds to the multiplicity of roots of identifying polynomial in Eq.~\ref{dimer-eq}. The degeneracy is lifted as the chemical potential deviates from the critical value, leading the mass parameter $\tilde{M}_F$ to deviate from zero. Consequently, away from the critical point the energy $\omega/\Delta$ varies with different slopes as the chemical potential varies and leads to the crossing of ferromagnetic and anti-ferromagnetic bound-state energies. Such crossings indicate transitions between the ferromagnetic and antiferromagnetic ground states of two magnetic impurities at specific chemical potentials. Presence of band inversion in bulk bands is then crucial for the realization of the latter transitions in the form of the YSR induced magnetic couplings. Fig. \ref{fig5}(b,d) represents the dimer energy for the non-topological phase ($m<0$) with different parameters. In the absence of the bulk band inversion, there is no crossing between two types of magnetic coupling in the non-topological phase.

\begin{figure*}[t]
\centering
\includegraphics[width=\textwidth]{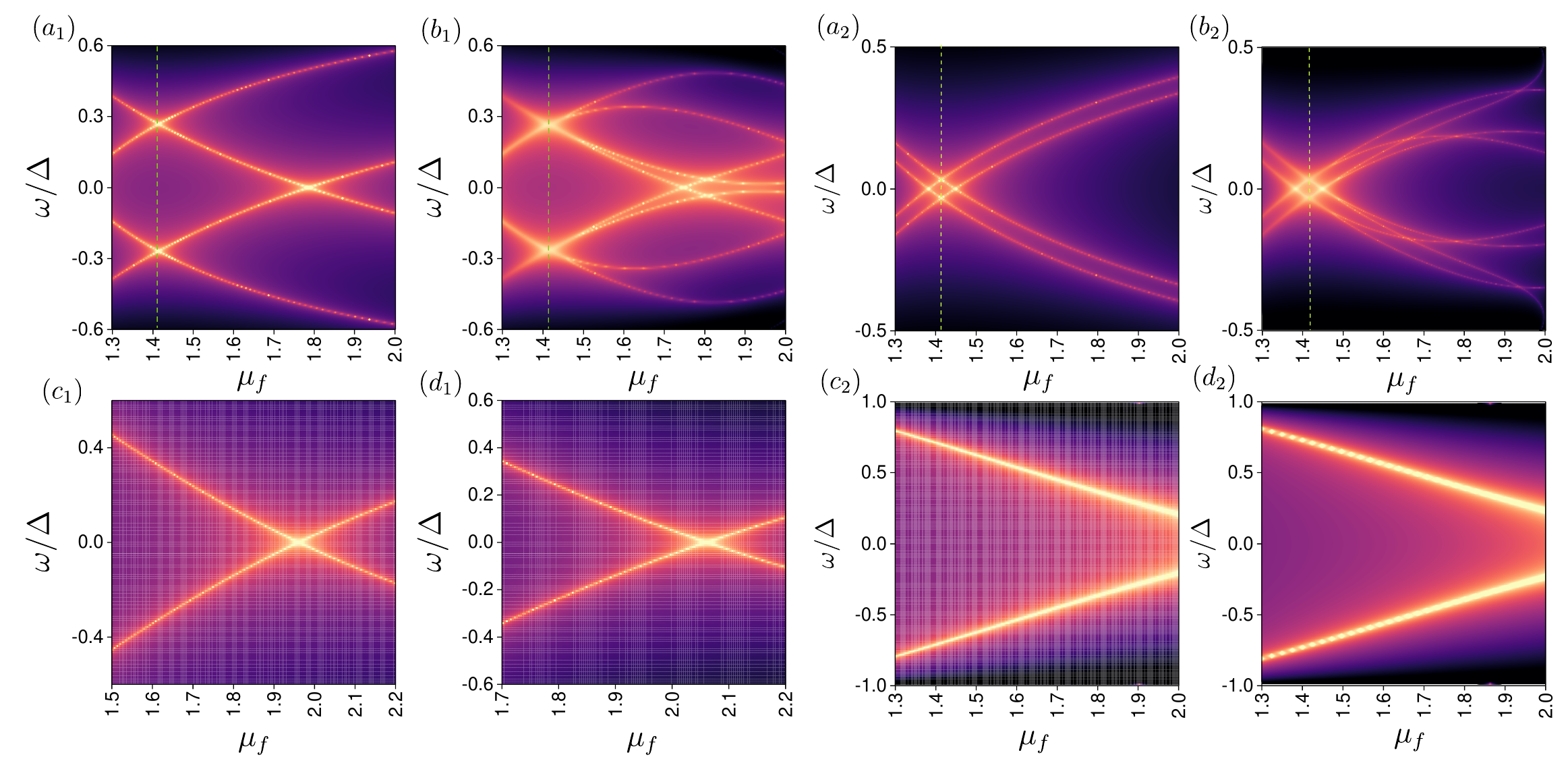} \caption{ The YSR states as a function of chemical potential induced by single impurity ($a_1,a_2$): 
in the bulk of a topological superconductor, and  ($b_1,b_2$): near its boundary. Their corresponding result for the non-topological cases are shown in ($c_1,c_2$): for the trivial bulk and ($d_1,d_2$): near the trivial boundary.
We set $d=3$, $M=1$, and $|m|=0.5$ (for topological (non-topological): $m>0$ ($m<0$)), and ${\cal J}S=26/\pi$ for ($a_1,b_1$), ${\cal J}S=20/\pi$ for ($a_2,b_2$). }
\label{fig7}
\end{figure*}
\section{Effect of the edge} The experimental signature of topological phases is commonly in the form of robust edge states that develop on the boundaries\cite{hasan2010colloquium,Xia2009}.When the bulk bands are doped and conducting, the edge states can scatter into the bulk but evidence has been presented that the signatures of edge states could still appear in the case of a superconducting doped topological insulator. An example in this regard is the presence of a zero-energy Majorana mode at the end of the vortex where they cross the edge of the sample \cite{PhysRevLett.107.097001}. It is demonstrated that this mode persists until the bulk doping reaches the critical doping level. It is then important to determine whether the distance of impurities from the edge could affect the YSR states in the superconducting phase. To this end, we need to develop a formalism to capture the boundary within the GF method. Previously, the effect of the hard boundary on the YSR states was captured by explicitly imposing the vanishing of the wave function at the boundary through modification of the green function \cite{GF1}. The latter construction, though, is not capable of capturing the effect of TI edge states when the bulk is doped. In this paper, we adopt a different approach that models the hard boundary by implementing the edge through an infinite potential in the GF method. The Schr\"{o}dinger equation in terms of the GF with impurity potential at the position $\textbf{r}_j$ and an infinite 2D impurity potential $V^{\text{HW}}=V \mu^3 \gamma^5$ placed at $z=d$ is written as
\begin{equation}
[\omega-H^{\text{BDG}}] \psi(\textbf{r})=\sum_j V_j^{\text{imp}} \delta(\textbf{r}-\textbf{r}_j) \psi(\textbf{r}_j)+V^{\text{HW}} \delta(z-d) \psi(\textbf{r}).
\end{equation}

The above Schr\"{o}dinger equation can be written in terms of GF with the hard boundary by taking the limit of $V \rightarrow +\infty$ in $V^{\text{HW}}$ [Appendix. \ref{appd}]:
\begin{equation} \label{BB}
\begin{split}
&\psi(0)=\sum_j V_j^{\text{imp}}  \lbrace G(\omega,0)\\ &- \int \dfrac{d \textbf{p}_\parallel}{(2\pi)^2} G(\textbf{p}_\parallel,-d)G(\textbf{p}_\parallel,0)^{-1}  G(\textbf{p}_\parallel,d) \rbrace \psi(0),
\end{split}
\end{equation}
where we consider $\textbf{r}_{1}=0$ for the single impurity case. 

The GF with hard boundary reduces to
\begin{equation} \label{GFB}
\begin{split}
&	G_{b}(\omega,d)= G_0(\omega)-\\ & \int \dfrac{d \textbf{p}_\parallel}{(2\pi)^2} G(\textbf{p}_\parallel,-d)G(\textbf{p}_\parallel,0)^{-1} G(\textbf{p}_\parallel,d).
\end{split}
\end{equation}
The derivation of equation \ref{GFB} is given in Appendix. \ref{appd}. In presence of the boundary, the YSR states with energy $\omega$ satisfies the following equation
\begin{equation}
\text{det}\lbrace 1-V^{\text{imp}}_1 G_b(\omega,d)\rbrace=0,
\end{equation}
where $d$ is the distance of the impurity from the boundary. The solution of the above equation represents boundary-induced modifications of YSR states as compared to the case where an impurity is located in the bulk (when $d \rightarrow \infty$, and $G_b(\omega,d)\rightarrow G_0(\omega)$). The results are shown in Fig. \ref{fig7} for both topological (top row with $m>0$) and non-topological (bottom row with $m<0$) cases. Additionally, a comparison of the YSR states is made between the situations in which an impurity is near the boundary (Fig. \ref{fig7}($b_1,b_2$)) and when it is in the bulk (Fig. \ref{fig7}($a_1,a_2$). Our findings demonstrate that when there is a band inversion in the bulk electronic bands, the modification of YSR states caused by the boundary can be significant. The main effect of the band is that, at certain chemical potential ranges, the boundary drives the YSR state to the lower energies. The primary effect of the edge would be the breaking of translation symmetry, which would lead to the formation of hybridization between YSR state orbitals. Such hybridization would push the energy of lower states toward zero energy.

The energy of YSR states depends not only on the distance away from the boundary but also explicitly depends on the strength of the impurity potential, as we can compare Fig. \ref{fig7}($b_1$) and ($b_2$). As can be seen from the results in the bottom row of Fig. \ref{fig7}, the YSR states in the non-topological case are significantly more boundary independent.
\\
\section{Conclusion} In this paper, we studied the effect of bulk band inversion on the YSR states in superconducting doped topological materials. Our results could potentially be realized in materials such as superconducting $Nb$-doped $Bi_2Se_3$ \cite{PhysRevLett.104.057001} and topological Fe-based superconductors \cite{Yin2015}. In addition to identifying novel phenomena in superconducting phases of materials with topologically non-trivial electronic states, our results could address some of puzzling features in the latter material\cite{zaki2021time,li2021electronic}. We should note that the phenomena explored in this paper result from a conventional type of intra-orbital pairing in the material. The novel property of the superconducting defect modes results from the bulk band inversion rather than unconventional SC pairing. The strong tunability of YSR states through the control of chemical potential points to the possibility of quantum phase transitions in magnetically doped topological materials which are tuned by doping level. In this paper we treated the impurity spin classically which is relevant for large spins. Studies of impurities with small spin, which should be analyzed quantum mechanically, as well as the possibilities of doping induced bulk magnetic phase transitions in the presence of dense magnetic impurities are the subjects of our future studies.

\section{acknowlegements}
We thank Rafael Fernandes and Ruixing Zhang for fruitful discussions. We acknowledge support from NSF HRD-2112550 (NSF CREST Center IDEALS) and NSF DMR-1824265 (DN and PG), NSF DMR-2130544 (PG and SH) and  NSF DMR-2047193 (PH).

\appendix
\section{Local Green's Function} \label{appa}
The superconducting GF corresponding to the Hamiltonian $H_{\text{BdG}}=\mu^3 (H(\textbf{p})-\mu_f \gamma^0)+\Delta \ \mu^2 \gamma^4$ where $		H(\textbf{p})= v_f \textbf{p}\cdot \gamma+ \tilde{M}(p) \gamma^5,$ is written as
\begin{widetext}
\begin{equation} \label{GF-Full}
G(\omega,p)= (\omega-H_{BdG})^{-1}=\dfrac{\mu^3 (\eta^2 (H-\mu_f)-(\mu_f^2-\lambda^2)(\mu_f+H))+\hat{\Delta} (\eta^2+(\mu_f+H)^2)}{(\eta^2+\epsilon_+^2)(\eta^2+\epsilon_-^2)}
\end{equation}
\end{widetext}
where $\epsilon_{\pm}=-\mu_f\pm \lambda$, $\lambda=\sqrt{v_f^2 p^2+\tilde{M}_F^2}$, $\eta=\sqrt{\Delta^2-\omega^2}$, $H=\tilde{M}_F \gamma^5+ \bf{p} \cdot \gamma$, and $\hat{\Delta}=\omega \mu^0+\Delta \mu^1$.
The full band GF in Eq. \ref{GF-Full} can be also written as a sum  ${\cal G}_++{\cal G}_-$ where
\begin{equation}
{\cal G}_{\pm}(\omega,\textbf{p})=\dfrac{1}{2 \lambda} \dfrac{\mu^3 \epsilon_{\pm}+\Delta \mu^1+\omega}{\epsilon_{\pm}^2+\Delta^2-\omega^2} (\lambda \gamma^0\pm \tilde{M}(p) \gamma^5 \pm  v_f \textbf{p}\cdot \gamma).
\end{equation}
The aforementioned ${\cal G}_\pm$ is equivalent to the contribution of the conduction and valence bands, respectively. They are valid for chemical potentials that cross the bands where $|\mu_F|> M$. The real-space GF at the origin ($r=0$) is given by
\begin{equation}
G_0(\omega)= \int_0^\infty \dfrac{d^3 p}{(2 \pi)^3} {\cal G}_{+} (\omega,p),
\end{equation}
leading to
\begin{equation}
G_0(\omega)=\dfrac{\nu_0}{2} \int \dfrac{\mu^3 \epsilon+\Delta \mu^1 +\omega}{\epsilon^2+\Delta^2-\omega^2} (\dfrac{\lambda \gamma^0+\tilde{M}(p) \gamma^5}{\lambda}) d\epsilon.
\end{equation}
Using the integrals $\int_{-\infty}^{+\infty} \dfrac{\epsilon}{\epsilon^2+\eta^2} d\epsilon=0$, $\int_{-\infty}^{+\infty} \dfrac{d\epsilon}{\epsilon^2+\eta^2}=\dfrac{\pi}{\eta}$, where $\eta=\sqrt{\Delta^2-\omega^2}$, and assuming the fact that the most contributions comes from the Fermi surface $p \sim p_f+\epsilon/v_f$, the local GF would be
\begin{equation}
G_0(\omega)= \dfrac{\pi \nu_0}{2 \eta}  (\omega+\Delta \mu^1)(\gamma^0+ \dfrac{\tilde{M}_F }{\mu_f}\gamma^5).
\end{equation}
\section{Non-local Green's Function}\label{appb}
The two-site GF propagator between spatial coordinates $r_i$ and $r_j$ is given by
\begin{equation}
G(\omega,r)=\int \dfrac{d\bf{p}}{(2\pi)^3} e^{i \bf{p}\cdot \bf{r}} G(\omega,\bf{p}),
\end{equation}
where $\bf{r}=\bf{r}_i-\bf{r}_j$.
Having assumed that the most contributions come from the FS, and $\boldsymbol{r}=  r \hat{z}$, the interacting GF is given by
\begin{equation}
\begin{split}
&G(\omega,r)=\int \dfrac{d^3p}{(2\pi)^3} \  e^{-i rp(\epsilon)\cos \theta} {\cal G}_+(\omega,p)= \\ &\dfrac{\nu_0}{2}\int_{-\Lambda}^{\Lambda}  d\epsilon \ d\cos \theta \ e^{-i rp(\epsilon) \cos \theta} \ {\cal G}_+(\omega,p),
\end{split}
\end{equation}
where 
\begin{equation}
{\cal G}_{+}(\omega,\textbf{p})=\dfrac{1}{2 \lambda} \dfrac{\mu^3 \epsilon_{+}+\Delta \mu^1+\omega}{\epsilon_{+}^2+\Delta^2-\omega^2} (\lambda \gamma^0+\tilde{M}(p) \gamma^5 + v_f \textbf{p}\cdot \gamma),
\end{equation}
is the conduction band GF. Using the following integrals
\begin{equation}
\int_{-\infty}^{+\infty} d\epsilon \ \dfrac{e^{i p(\epsilon) r \cos\theta}}{\epsilon^2+\eta^2} =\dfrac{\pi}{\eta} \exp(ip_f r \cos\theta-\dfrac{\eta}{v_f}|r \cos\theta|)
\end{equation}
and 
\begin{equation}
\begin{split}
&\int_{-\infty}^{+\infty} d\epsilon \ \dfrac{\epsilon \ e^{i (p_f+\epsilon/v_f) r \cos\theta}}{\epsilon^2+\eta^2}\\ & =i \pi \ sign(r \cos\theta) \exp(ip_f r \cos\theta-\dfrac{\eta}{v_f}|r \cos\theta|),
\end{split}
\end{equation}
the interacting GF is would be
\begin{equation} 
\begin{split}
& G(\omega,r)=\\ &\dfrac{\pi \nu_0}{2} [ ( \dfrac{\hat{\Delta}}{\eta} f_1(q)+\mu^3 f_2(q)) (\gamma_0+\dfrac{\tilde{M}_F}{\mu_f} \gamma^3)\\ & +i \lbrace (\dfrac{\hat{\Delta} \ }{\eta}\dfrac{ v_f p_f}{\mu_f}-\dfrac{\eta}{\mu_f} \mu^3) f_3(q)+  (\mu^3  \dfrac{ v_f p_f}{\mu_f}+\dfrac{\hat{\Delta}}{\mu_f})f_4 (q)\rbrace \gamma^3 ],
\end{split}
\end{equation}
where we define $q=r p_f$, and
\begin{equation}
\begin{split}
& f_1(q)=\dfrac{\sin q}{q},  
\ \ f_2(q)=\dfrac{\cos q -1}{q},\\ &
f_3(q)=\dfrac{\sin q-q \cos q}{q^2},  
\ \ f_4(q)=\dfrac{q \sin q+ \cos q-1}{q^2}.
\end{split}
\end{equation}
We assumed $e^{-r/\xi_c} \approx 1$ since $r\ll \xi_c$, in which $\xi_c=v_f/\Delta$	is the coherence length of superconductor.
\\
\section{YSR Wave Function}\label{appc}
The wave function of YSR states at the position of $\bf{r}$ is defined as
\begin{equation}
\begin{split}
&\Psi(\textbf{r})=\int \dfrac{d \textbf{p}}{(2\pi)^3} e^{i \textbf{p}\cdot \textbf{r}} \psi(\textbf{p}) \\ &=\dfrac{\mp {\cal J}S \nu_0}{2}\int d \epsilon_p \ d\cos\theta \ e^{i p r \cos(\theta-\alpha)}  {\cal G}_+(\omega,\textbf{p}) \psi(0),
\end{split}
\end{equation}
where $\psi(0)$ is the wave function at the position of the impurity, and $\sphericalangle (\bf{r},\bf{p})=\theta-\alpha$. The $\textbf{r}$-dependent wave function is then given by
$\Psi(\textbf{r})=\dfrac{1}{\sqrt{N}}\lbrace \psi_1^+ \ket{\uparrow},\psi_1^- \ket{\uparrow}, \psi_2^+\ket{\uparrow},\psi_2^- \ket{\uparrow} \rbrace $
where $\ket{\uparrow}=(1,0)$, and $N=\int d \textbf{r} |\psi(\textbf{r})|^2$. The components of wave function are given by
\begin{equation} \label{bulk-comp}
\begin{split}
\psi_1^{\pm}=&\beta_{\pm} (\dfrac{\omega^\prime}{\eta} f_1(q,\alpha)-f_2(q,\alpha))+\\ &i \tilde{\nu} \dfrac{v_f p_f}{\mu_f}  (\dfrac{\omega^\prime}{\eta} f_3(q,\alpha)+f_4(q,\alpha)), \\
\psi_2^{\pm}=&\beta_{\pm} (\dfrac{\omega^\prime}{\eta} f_1(q,\alpha)+f_2(q,\alpha)+\\ &i \tilde{\nu} \dfrac{v_f p_f}{\mu_f}  (\dfrac{\omega^\prime}{\eta} f_3(q,\alpha)-f_4(q,\alpha)),
\end{split}
\end{equation}
where  $\omega^\prime=\omega+\Delta$, and 
\begin{equation}
\begin{split}
& f_1(q,\alpha)=\dfrac{1}{2} \int_{-1}^{1} e^{iq \cos(\theta-\alpha)} d\cos \theta, \\ & f_2(q,\alpha)=\dfrac{1}{2} \int_{-1}^{1} \text{sign}(r \cos \theta) \ e^{iq \cos(\theta-\alpha)} d\cos \theta \\
& f_3(q,\alpha)=\dfrac{1}{2} \int_{-1}^{1} \cos \theta \ e^{iq \cos(\theta-\alpha)} d\cos \theta , \\ & f_4(q,\alpha)=\dfrac{1}{2} \int_{-1}^{1} \text{sign}(r \cos \theta) \cos \theta \ e^{iq \cos(\theta-\alpha)} d\cos \theta.
\end{split}
\end{equation}
\section{Green's Function in the presence of hard boundary} \label{appd}
The Schr\"{o}dinger equation can be written in terms of the GF with impurity potentials at the RHS. We consider multiple impurities placed at the positions $r_j$, and an infinite 2D impurity potential $V^{\text{HW}}=V \mu^3 \gamma^5$ placed at $z=d$ acting like the hard wall
\begin{equation}
[\omega-H^{\text{BDG}}] \psi(\textbf{r})=\sum_j V_j^{\text{imp}} \delta(\textbf{r}-\textbf{r}_j) \psi(\textbf{r}_j)+V^{\text{HW}} \delta(z-d) \psi(\textbf{r}).
\end{equation}
Using the identity $\psi(\textbf{r})=\int \dfrac{d \textbf{p}}{(2\pi)^3} e^{i \textbf{p}\cdot \textbf{r}} \psi(\textbf{p})$, and
multiplying $\int e^{-i\textbf{p}^\prime\cdot \textbf{r}} d\textbf{r}$ from the left and then multiplying $\int e^{-i\textbf{p}^\prime\cdot \textbf{r}_i} \dfrac{d \textbf{p}^\prime}{(2\pi)^3}$ to both sides, the above equation takes the form
\begin{equation} \label{WFb}
\begin{split}
\psi(\textbf{r}_i)=&\sum_j V_j^{\text{imp}} G(\omega,\textbf{r}_{ij}) \psi(\textbf{r}_j)+\\ & \int \dfrac{d\textbf{p}_\parallel}{(2\pi)^2} e^{i \textbf{p}_\parallel \cdot \textbf{r}_{\parallel,i}}G(\textbf{p}_\parallel,-d) V^{\text{HW}} \psi(\textbf{p}_\parallel,d),
\end{split}
\end{equation}
where we assume the impurity is placed at the origin and the boundary is at the distance of $d$. Using the previous procedures, the wave function $\psi(\textbf{p}_\parallel,d)$  can be obtained as
\begin{equation}
\begin{split}
&\psi(\textbf{p}_\parallel,d)=\\ &G(\textbf{p}_\parallel,d) (1-V^{\text{HW}} G(\textbf{p}_\parallel,0))^{-1}\sum_j V_j^{\text{imp}} \psi(\textbf{r}_j) e^{-i \textbf{p}_\parallel\cdot \textbf{r}_{\parallel,j}}.
\end{split}
\end{equation}

Plugging $\psi(\textbf{p}_\parallel,d)$ into Eq.~\ref{WFb}, and taking the limit of $V \rightarrow +\infty$, we will get
\begin{equation} \label{BB}
\begin{split}
&\psi(\textbf{r}_i)=\sum_j V_j^{\text{imp}} \lbrace G(\omega,\textbf{r}_{ij})\\ &- \int \dfrac{d \textbf{p}_\parallel}{(2\pi)^2} e^{-i \textbf{p}_\parallel \cdot \textbf{r}_{\parallel,ij}} G(\textbf{p}_\parallel,-d)G(\textbf{p}_\parallel,0)^{-1}  G(\textbf{p}_\parallel,d) \rbrace \psi(\textbf{r}_j) ,
\end{split}
\end{equation}
where $\textbf{r}_{ij}=0$ for the single impurity. The boundary GF, therefore, reduces to
\begin{equation} \label{GFB}
\begin{split}
&	G_{b}(\omega,d)= G_0(\omega)-\\ & \int \dfrac{d \textbf{p}_\parallel}{(2\pi)^2} G(\textbf{p}_\parallel,-d)G(\textbf{p}_\parallel,0)^{-1} G(\textbf{p}_\parallel,d).
\end{split}
\end{equation}

Using the identity $G(\textbf{p}_\parallel,d)=\int \dfrac{dp_z}{2\pi} e^{i p_z d} G (\textbf{p})$, and change the integration over $p_z$ into the integration over energy $\epsilon$, i.e., $\frac{dp_z}{2\pi}=F(\epsilon,\textbf{p}_\parallel) d\epsilon$, with fixed $\textbf{p}_\parallel$, we get
\begin{equation}
\int \dfrac{dp_z}{2 \pi} e^{i p_z d} G(\textbf{p})=\int F(\epsilon,\textbf{p}_\parallel) e^{-ip_z(\epsilon,\textbf{p}_\parallel)d} G(\epsilon,\textbf{p}) d\epsilon
\end{equation}
where 
\begin{equation}
F(\epsilon,\textbf{p}_\parallel)=\dfrac{\epsilon+\mu_f}{2 \pi v_f^2 p_z(\epsilon,\textbf{p}_\parallel)(1-2m \tilde{M}(\epsilon,\textbf{p}_\parallel))}.
\end{equation}
Given the weak pairing type of superconductivity considered, the dominant contributions to the in-gap states emerge from electronic states with energy close to the chemical potential. As a result, we can keep the terms up to the first order of $\epsilon$

\begin{equation}
\begin{split}
& F(\epsilon,\textbf{p}_\parallel)=F^{(0)}(\mu_f,\textbf{p}_\parallel)+F^{(1)} (\mu_f,\textbf{p}_\parallel) \epsilon +{\cal O}(\epsilon^2), \\
& p_z(\epsilon,\textbf{p}_\parallel)=p_z^{(0)} (\mu_f,\textbf{p}_\parallel)+p_z^{(1)} (\mu_f,\textbf{p}_\parallel) \epsilon+{\cal O}(\epsilon^2).
\end{split}
\end{equation}
which leads to the following integration
\begin{widetext}
\begin{equation}
\begin{split}
	&\int F(\epsilon,\textbf{p}_\parallel) e^{-i p_z(\epsilon,\textbf{p}_\parallel)d} {\cal G}(\epsilon,\textbf{p}) d\epsilon=
	\\ & \dfrac{1}{2} \lbrace F^{(0)}(\mu_f,\textbf{p}_\parallel) (\gamma^0+\dfrac{\tilde{M}_F}{\mu_f}\gamma^5+\dfrac{v_f \textbf{p}_\parallel \cdot \gamma}{\mu_f}) +\dfrac{v_f p_z^{(0)} }{\mu_f} F^{(0)}(\mu_f,\textbf{p}_\parallel) \gamma^3 \rbrace e^{-i p_z^{(0)}(\textbf{p}_\parallel)d} \int  e^{-i p_z^{(1)}(\textbf{p}_\parallel) \epsilon d} \ \dfrac{\mu^3 \epsilon + \hat{\Delta}}{\epsilon^2+\eta^2} \ d \epsilon 
	\\ & +\dfrac{1}{2} \lbrace F^{(1)}(\mu_f,\textbf{p}_\parallel) (\gamma^0+\dfrac{\tilde{M}_F}{\mu_f}\gamma^5+\dfrac{v_f \textbf{p}_\parallel \cdot \gamma}{\mu_f}) +\dfrac{v_f p_z^{(0)}(\textbf{p}_\parallel)}{\mu_f} \gamma^3 \rbrace e^{-i p_z^{(0)}(\textbf{p}_\parallel)d} \int  e^{-i p_z^{(1)}(\textbf{p}_\parallel) \epsilon d} \ \dfrac{\epsilon \  \hat{\Delta}}{\epsilon^2+\eta^2} d \epsilon.
\end{split}
\end{equation}

Using the identities
\begin{equation}
\begin{split}
	& \int \dfrac{\epsilon}{\epsilon^2+\eta^2} e^{-i p_z^{(1)} (\textbf{p}_\parallel)\epsilon d} d\epsilon=-i \pi e^{-\eta|p_z^{(1)}(\textbf{p}_\parallel)d|} \text{sign}(p_z^{(1)}(\textbf{p}_\parallel) d), \\
	& \int \dfrac{1}{\epsilon^2+\eta^2} e^{-i p_z^{(1)} (\textbf{p}_\parallel)\epsilon d} d\epsilon=\dfrac{\pi}{\eta} e^{-\eta|p_z^{(1)}(\textbf{p}_\parallel)d|},
\end{split}
\end{equation}

the boundary GF, $G(\textbf{p}_\parallel,d)$, would be
\begin{equation}
\begin{split}
	& G(\textbf{p}_\parallel,d)= \\
	&	\lbrace \dfrac{\pi}{2} (-i \mu^3 \text{sign}(p_z^{(1)}(\textbf{p}_\parallel) d)+\dfrac{\hat{\Delta}}{\eta})( F^{(0)}(\mu_f,\textbf{p}_\parallel) (\gamma^0+\dfrac{\tilde{M}_F}{\mu_f}\gamma^5+\dfrac{v_f \textbf{p}_\parallel \cdot \gamma}{\mu_f}) +\dfrac{v_f p_z^{(0)} }{\mu_f} F^{(0)}(\mu_f,\textbf{p}_\parallel) \gamma^3 ),
	\\ & +\dfrac{\pi}{2} (-i \hat{\Delta} \ \text{sign}(p_z^{(1)}(\textbf{p}_\parallel) d) ) ( F^{(1)}(\mu_f,\textbf{p}_\parallel) (\gamma^0+\dfrac{\tilde{M}_F}{\mu_f}\gamma^5+\dfrac{v_f \textbf{p}_\parallel \cdot \gamma}{\mu_f}) +\dfrac{v_f p_z^{(0)}(\textbf{p}_\parallel)}{\mu_f} \gamma^3)  \rbrace e^{-i p_z^{(0)}(\textbf{p}_\parallel)d} e^{-\eta|p_z^{(1)}(\textbf{p}_\parallel)d|} .
\end{split}
\end{equation}
In above equations, we defined  $\hat{\Delta}=\omega\mu^0+\Delta \mu^1$, $\gamma^0=\tau^0 \otimes \sigma^0$, $\gamma^1=\tau^1 \otimes \sigma^1$, $\gamma^2=\tau^1 \otimes \sigma^2$, $\gamma^3=\tau^1 \otimes \sigma^3$, $\gamma^5=\tau^3 \otimes \sigma^0$, and $\gamma=(\gamma^1,\gamma^2,\gamma^3)$, where the Pauli matrices $\sigma^i$ and $\tau^i$ act on the spin and orbital degrees of freedom, respectively.
\end{widetext}

\bibliography{ref}

\end{document}